\newcommand{\Ox}{\(\mathrm{^{16}O}\)\xspace}
\newcommand{\reac}{\(\mathrm{^{16}O}(p,\alpha)\mathrm{^{13}N}\)\xspace}
\newcommand{\N}{\(\mathrm{^{13}N}\)\xspace}
\newcommand{\F}{\(\mathrm{^{17}F}\)\xspace}
\newcommand{\reacE}{\(\mathrm{^{13}N}(\alpha,p)\mathrm{^{16}O}\)\xspace}
\newcommand{\event}{\((p,\alpha)\)\xspace}

\newcommand{\figf}[1]{Figure~\ref{#1}\xspace}

\documentclass[twocolumn,trackchanges]{aastex701}
\usepackage{graphicx}
\usepackage{subcaption}

\usepackage{xspace}
\shorttitle{Rate of $^{16}$O$(p,\alpha)^{13}$N from MUSIC}

\begin{document}


\title{
Implications for Type 1a supernovae nucleosynthesis from an \\
experimentally constrained $^{16}$O$(p,\alpha)^{13}$N reaction rate
}

\author[orcid=0000-0002-1238-3059]{M. Alruwaili}
\affiliation{ School of Physics, Engineering and Technology, University of York, York YO10 5DD, UK}
\affiliation{Physics Department, Faculty of Science, Northern Border University, Arar 13211, Saudi Arabia}
\email[show]{ma1722@york.ac.uk}

\author[orcid=0000-0002-1236-4739]{C. Foug\`eres}
\affiliation{ Physics Division, Argonne National Laboratory, Lemont, IL 60439, USA}
\email{chloe.fougeres@cea.fr}

\author[orcid=0000-0003-0423-363X]{A.M. Laird}
\affiliation{ School of Physics, Engineering and Technology, University of York, York YO10 5DD, UK}
\email{alison.laird@york.ac.uk}  

\author[orcid=0000-0001-8746-0234]{H. Jayatissa}
\affiliation{ Physics Division, Argonne National Laboratory, Lemont, IL 60439, USA}
\email{hesh@lanl.gov}  
\author [orcid=0009-0002-4051-9627]{ M. L. Avila}
\affiliation{ Physics Division, Argonne National Laboratory, Lemont, IL 60439, USA}
\email{mavila@anl.gov} 

\author[orcid=0000-0003-0894-6450]{E. Bravo}
\affiliation{Departamento de Física Teórica y del Cosmos, Universidad de Granada, 18071 Granada, Spain}
\email{ eduardo.bravo@upc.edu} 
\author[orcid=0000-0003-0778-4429]{C. Angus}
\affiliation{ School of Physics, Engineering and Technology, University of York, York YO10 5DD, UK}
\affiliation{ TRIUMF, 4004 Wesbrook Mall, Vancouver V6T 2A3, BC, Canada}
\email{cangus@triumf.ca} 

\author[orcid=0000-0003-3494-343X]{C. Badenes}
\affiliation{Department of Physics and Astronomy, University of Pittsburgh, 3941 O’Hara Street, Pittsburgh, PA 15260, USA}
\affiliation{Pittsburgh Particle Physics, Astrophysics, and Cosmology Center (PITT PACC), University of Pittsburgh, Pittsburgh, PA 15260, USA}
\email{ badenes@pitt.edu} 

\author[orcid=0000-0003-3804-460X]{S. Chakraborty}
\affiliation{ School of Physics, Engineering and Technology, University of York, York YO10 5DD, UK}
\email{soham.chakraborty@york.ac.uk}  

\author[orcid=0000-0002-9778-8759]{C. Diget}
\affiliation{ School of Physics, Engineering and Technology, University of York, York YO10 5DD, UK}
\email{christian.diget@york.ac.uk}  

\author[orcid=0000-0002-6150-2085]{N. de Séréville}
\affiliation{Institut de Physique Nucléaire d’Orsay, UMR8608, IN2P3-CNRS, Université Paris Sud 11, 91406 Orsay, France}
\email{deserevi@ipno.in2p3.fr}

\author[orcid=0000-0002-6734-6204]{A. Hall-Smith}
\affiliation{ School of Physics, Engineering and Technology, University of York, York YO10 5DD, UK}
\email{alexander.hall-smith@york.ac.uk}  

\author[orcid=0000-0001-7731-580X]{R. Longland}
\affiliation{Department of Physics, North Carolina State University, Raleigh, NC 27695, USA}
\affiliation{
Triangle Universities Nuclear Laboratory, Durham, NC 27708, USA}
\email{richard_longland@ncsu.edu}

\author{W.-J. Ong}
\affiliation{Lawrence Livermore National Laboratory, Livermore, CA 94550, USA}
\email{ong10@llnl.gov}

\author{K. E. Rehm}
\affiliation{ Physics Division, Argonne National Laboratory, Lemont, IL 60439, USA}
\email{rehm@anl.gov} 

\author[orcid=0000-0003-3125-9907]{D. Santiago-Gonzalez}
\affiliation{ Physics Division, Argonne National Laboratory, Lemont, IL 60439, USA}
\email{dasago@anl.gov}

\begin{abstract}
The \(\mathrm{^{16}O}(p,\alpha)\mathrm{^{13}N}\) reaction plays a key role in shaping the $\alpha$-particle abundance during explosive oxygen burning in Type Ia supernovae. By enhancing $\alpha$-production, this reaction directly affects the calcium-to-sulphur (Ca/S) and argon-to-sulphur (Ar/S) ratios, which serves as a tracer of progenitor metallicity. However, recent work suggests that the rate must be enhanced by a factor of up to seven over the standard value to explain observed Ca/S ratios across a range of progenitor metallicities. To explore this impact, available experimental cross-section data for the \(\mathrm{^{16}O}(p,\alpha)\mathrm{^{13}N}\) reaction have been compiled and critically evaluated. Significant discrepancies are identified in the low-energy region (\( E_{\mathrm{cm}} \) = 5.7–7.0~MeV), primarily due to limitations of the activation method. To resolve this, the first direct measurement at astrophysical energies has been performed using the MUSIC active-target detector. The new \(\mathrm{^{16}O}(p,\alpha)\mathrm{^{13}N}\) thermonuclear reaction rate is found to be approximately 1.5 times higher than the REACLIB rate in the temperature range T = 3–4 GK, with more constrained uncertainties that resolve the previously large spread among existing data. 
The suggested factor of seven enhancement is excluded and these results indicate that this reaction alone cannot fully explain the variation in the Ca/S and Ar/S ratios observed across different progenitor metallicities. 
Therefore, future work should focus on 
reducing the uncertainties in other key oxygen-burning reactions, particularly \(\mathrm{^{16}O+^{16}O}\) and \(\mathrm{^{12}C+^{16}O}\).
Further reducing the constraints on the
\(\mathrm{^{16}O}(p,\alpha)\mathrm{^{13}N}\) rate is also needed to fully determine to whether a nuclear physics solution to this discrepancy is possible.
\end{abstract}


\keywords{\uat{Nuclear astrophysics}{1129} --- 
          \uat{Nuclear reaction cross sections}{2087} --- 
          \uat{Type Ia supernovae}{1728} --- 
          \uat{Nucleosynthesis}{1131} --- 
          \uat{Reaction rates}{2081}}


\section{Introduction}
\label{sec:level1}

Type Ia supernovae (SNe Ia) play a crucial role in the chemical evolution of the galaxy. In addition to iron-group elements, they are known to synthesise significant amounts of intermediate-mass nuclei, including silicon, sulphur, and calcium~\citep{Leung2020Explosive,Thielemann2018Nucleosynthesis,jose2016stellar}. These thermonuclear explosions occur in carbon-oxygen white dwarfs, and their spectra are characterised by strong silicon features at the epoch of maximum brightness~\citep{Mazzali2007A}. When temperatures rise above 3~GK, oxygen-burning reactions are activated~\citep{Zhao2016THE,woosley1973explosive}. Among these, the \reac reaction plays a key role by enhancing $\alpha$-particle abundance. Protons are released by oxygen-burning reactions, such as $^{16}$O($^{16}$O,p)$^{31}P$. Then $^{16}$O is converted into $^{12}$C through the chain $^{16}$O$(p,\alpha)^{13}$N$(\gamma,p)^{12}$C, releasing $\alpha$-particles. As a result, the effective $\alpha$-production per oxygen nucleus destroyed is significantly boosted. Increased $\alpha$-production favours the synthesis of calcium over sulphur, thereby influencing the calcium-to-sulphur Ca/S production ratio~\citep{woosley1972astrophysical,bravo2012sensitivity}.

Observational evidence indicates that the Ca/S ratio in SNe Ia ejecta varies with progenitor metallicity~\citep{martinez2017observational,De2014,bravo201916o}, suggesting a sensitivity to the \reac reaction rate. 
Based on a modelling study constrained by supernova remnants observations, \citet{bravo201916o} proposed that a rate enhancement of up to a factor of seven over the CF88 compilation~\citep{Caughlan1988Thermonuclear} is required to explain the full range of Ca/S observed at different metallicities.

However, the reaction rate for \reac remains subject to significant uncertainties.~\citet{wagoner1969synthesis}~introduced the first rate, although the numerical basis for this rate was not explicitly detailed. Later,~\citet{woosley1973explosive}~pointed out that Wagoner's rate likely overestimated the contribution of the \reac reaction during explosive nucleosynthesis.

The REACLIB database~\citep{cyburt2010jina} adopted the rate from the CF88 compilation~\citep{Caughlan1988Thermonuclear}, which remains widely used in stellar modelling. However, no associated uncertainties are provided.

More recently, STARLIB~\citep{sallaska2013starlib} applied a Hauser–Feshbach (H-F) statistical model to estimate the reaction rate and its uncertainty bands. However, the H-F approach relies on averaged nuclear properties and is generally less reliable for systems with low-level density, such as the compound nucleus $^{17}$F. STARLIB therefore assigns a factor of ten uncertainty in the rate~\citep{sallaska2013starlib}.

A review of existing experimental cross section data shows that although the measurements are broadly consistent above $E_{\mathrm{p}} = 7.4$~MeV ($ \approx E_{\mathrm{cm}} = 7.0$~MeV), significant discrepancies remain at lower energies, particularly in the energy range $E_{\mathrm{p}} = 6.0$-7.4~MeV ($\approx E_{\mathrm{cm}} = 5.6$–7.0~MeV) (Figure~\ref{fig:exfordata}). In this region, cross sections differ by more than reported uncertainties, limiting the ability to reliably constrain the \reac reaction rate at the temperatures relevant to explosive oxygen burning in SNe Ia. 

The goal of this work is to resolve these discrepancies through a new direct measurement of the \reac cross section using the MUlti Sampling Ionisation Chamber (MUSIC) active-target detector~\citep{Carnelli2015}. The new data provide improved coverage at low centre-of-mass energies and are used to derive a revised thermonuclear reaction rate. The implications of the updated rate for stellar models and Ca/S production in SNe Ia are also discussed.

The structure of this paper is as follows: 
Section~\ref{sec:previous_data} presents prior experimental data and their inconsistencies. Section~\ref{sec:music_measurement} introduces the new measurement performed with the MUSIC detector. The result of the cross section calculation is summarised in Section~\ref{sec:Crossdata}. In Section~\ref{sec:NewRate}, the updated reaction rate is shown and discussed. Finally, Section~\ref{sec:RateImpact} discusses the astrophysical implications of the new rate. The conclusions of this work are presented in Section~\ref{sec:con}.

\section{Limitations on The Existing \reac cross section Data}
\label{sec:previous_data}
\begin{figure}[htbp]
    \centering
    \includegraphics[width=1\linewidth]{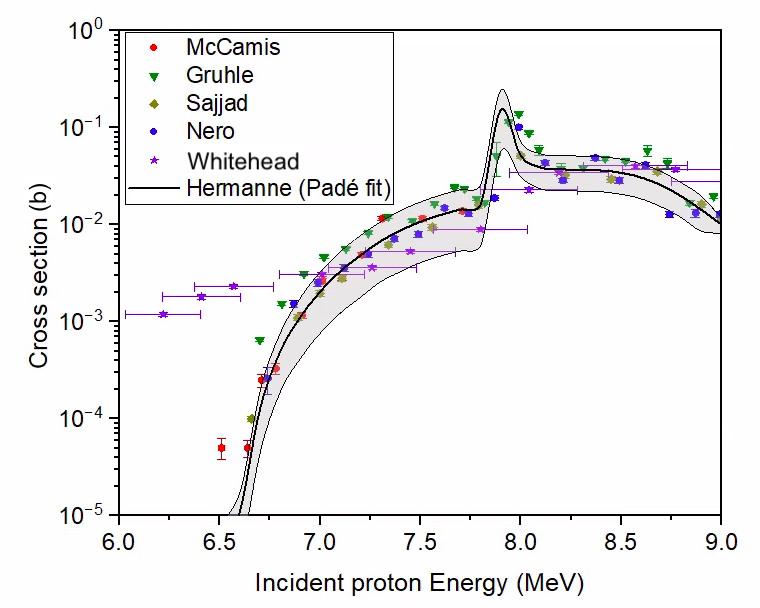}
     \caption{Experimental cross section data for the \reac\ reaction, plotted as a function of incident proton energy in the laboratory frame. The data were extracted from the EXFOR database~\citep{zerkin2018experimental} and include five datasets: ~\citet{mccamis1973total}, ~\citet{gruhle1977reactions},~\citet{SAJJAD1986Cyclotron}, ~\citet{NERO}, and ~\citet{whitehead1958activation}. The solid line represents the recommended cross section evaluation using Padé approximant fits from~\citet{Hermanne2021}.}

    \label{fig:exfordata}
\end{figure}

Several experimental datasets for the \reac\ cross section have been reported over the past decades, including measurements by \citet{whitehead1958activation}, \citet{NERO}, \citet{mccamis1973total}, \citet{gruhle1977reactions}, and \citet{SAJJAD1986Cyclotron}.
A direct comparison of these datasets, extracted from the EXFOR database \citep{zerkin2018experimental} (see Figure~\ref{fig:exfordata}),
 reveals significant inconsistencies in both the absolute cross section values and the shape of the excitation function. The largest discrepancies occur in $E_{\mathrm{p}} = 6$–$7.5$~MeV ($E_{\mathrm{cm}} = 5.7$–$7.1$~MeV)  where cross sections differ by up to a factor of 2–3 at the same energy, exceeding the stated experimental uncertainties.

It should be noted that the Whitehead dataset, as reported in EXFOR \citep{zerkin2018experimental}, includes a remark stating that the data were uniformly shifted by 0.3~MeV to lower energies by the NNDC~\citet{EXFOR2Whit}. This shift was applied without an explicit scientific justification. In addition, the data showed unrealistically large cross sections near the reaction threshold and presented with no clear uncertainty on the cross section. As a result, the Whitehead dataset was excluded from the subsequent analysis.

Later, \citet{Meyer2020EvaluationGrains} evaluated the $^{17}$F level information above the $\alpha$-particle threshold, focusing primarily on extracting $\alpha$-particle partial widths. The deduced parameters were then used as input to the \texttt{AZURE2} $\mathcal{R}$-matrix code \citep{azuma2010azure} to calculate the cross section for the $^{13}$N$(\alpha,p)^{16}$O reaction over the energy range $E_{\mathrm{cm}}({^{13}\mathrm{N}+\alpha}) = 0$–2.5~MeV, corresponding to $E_{\mathrm{cm}}({^{16}\mathrm{O}+p}) = 5.2$–7.7~MeV for the inverse \reac reaction. However, we note the following. First, significant uncertainties remain in the properties of the $^{17}$F states themselves, particularly in the assumptions made regarding unknown $\alpha$-particle widths. Second, although they have discussed the possible  interference  between resonances, these interference effects were neglected due to their limited impact on the rate at the lower temperatures considered in that work. 

More recently, \citet{Hermanne2021} performed an updated evaluation of positron emitter production reactions, including the \reac. This work updated the earlier evaluation by \citet{takacs2003validation}, and employed Padé approximant fits \citep{pade1892representation} with 40 parameters to selected experimental cross section datasets in order to deduce recommended values with associated uncertainties.

While the Padé fit by~\citet{Hermanne2021} reproduces the general trend of the experimental data at higher energies, its applicability at lower energies ($E_{\mathrm{cm}} \approx 5.5$–7.0~MeV) is more limited. In this region, strong variations between datasets, aggravated by potential systematic issues such as the energy shift observed in the Whitehead data introduce significant uncertainties. As a result, the IAEA~\citep{Hermanne2021} evaluation carries an uncertainty of approximately 65\% for the lower energy cross-section, further illustrating the lack of robust experimental constraints in this energy region.

Earlier studies of \reac employed the activation method, which requires irradiating a target with the beam, then transferring the target to a separate detection system to measure the delayed \(\beta^+\) decay of the \N product nuclei. This indirect approach introduces several systematic uncertainties: potential loss of \N nuclei during the transfer process, background contributions from other radioactive products, and difficulties in precisely determining the effective target thickness and beam integration over the irradiation period.

In contrast, the present measurement using the MUSIC detector directly identified \reac reaction events in real-time. The active-target approach employed in MUSIC  \citep[for more details on MUSIC structure, see][]{Carnelli2015} eliminates these complications by detecting the reaction products directly at the moment of their creation. This direct detection method is less susceptible to the backgrounds that contribute to activation measurements.

\section{Experimental Setup and MUSIC Measurement}
\label{sec:music_measurement}
To address the discrepancies in the low-energy cross section measurements of \reac, a new experiment was carried out at the Argonne Tandem Linac Accelerator System (ATLAS) at Argonne National Laboratory. 

MUSIC is an active-target detector in which the filling gas serves simultaneously as both the reaction target and the detection medium. As the beam travels through the gas, it loses energy at a rate that depends on its charge and mass. When a reaction occurs, the rate of energy loss changes and so the reaction products can be identified by this change in energy loss. Inside the MUSIC detector, the anode is divided into 18 strips, each 15.78~mm wide, arranged along the beam axis. This segmentation enables tracking of the gradual energy loss as the beam propagates through the gas. Consequently, reactions occur at different positions inside the detector, corresponding to a range of centre-of-mass energies for a single beam energy. The 16 central strips (1–16), known as active anodes, are divided into two sections labelled (L) and (R).

The measurement of the \reac\ reaction was performed in inverse kinematics with a $^{16}$O beam at $135.5 \pm 0.15\,\mathrm{MeV}$. The MUSIC chamber was filled with pure methane gas ($\mathrm{CH}_4$) at a pressure of $710 \pm 1\,\mathrm{~Torr}$. After passing through a 1.3\,mg/cm$^2$ titanium entrance window and a 35.9\,mm dead-gas region, the beam energy at the entrance to the active detection volume was about 123.9~MeV.

The beam intensity was maintained at approximately $3\times10^{4}$ particles per second to maintain stable data acquisition. The experiment was conducted over 27 hours, with data collection and monitoring performed using the CoMPASS software\cite{CAENu}.

The centre-of-mass energy corresponding to each strip depends on the cumulative energy loss of the $^{16}$O beam while traversing the gas target. The actual energy loss of the $^{16}$O beam particles at different gas pressures was measured using a silicon detector mounted outside the chamber. These silicon-detector measurements were compared with theoretical calculations from ATIMA and SRIM. The ATIMA calculations showed better agreement than with SRIM. Therefore, the centre-of-mass energy corresponding to each strip, along with its associated uncertainty, was determined using Monte Carlo simulations based on ATIMA stopping-power tables \citep{ATIMA}. The uncertainty in the centre-of-mass energy was driven primarily by the strip width (15.78\,mm) and the beam energy loss within each strip.

In the present setup, $(p,\alpha)$ reaction events were identified by a characteristic decrease in energy loss ($\Delta E$) relative to the beam signal, because $^{13}$N has a lower atomic number than $^{16}$O. With a Q-value of –5.218~MeV for the \reac\ reaction, the beam energy drops below the reaction threshold by strip 6. The \reac\ reaction events were therefore confined to strips 2–6, corresponding to centre-of-mass energies above the reaction threshold. Both $^{13}$N and $^{16}$O ions were fully stopped within the detector volume. 
To improve the visibility of the $(p,\alpha)$ energy loss pattern, the $^{16}$O beam traces were normalised across all strips to an arbitrary value of 6 (a.u.). Normalisation removed the Bragg peak but improved the contrast between the beam and the reaction products. As the $^{16}$O beam traversed the active volume, it lost 4–5~MeV of energy per segmented anode strip. This corresponded to a centre-of-mass energy range from 6.9~MeV down to 5.8~MeV across strips 2–6.

\begin{figure}[ht]
\centering
\includegraphics[width=1.0\linewidth]{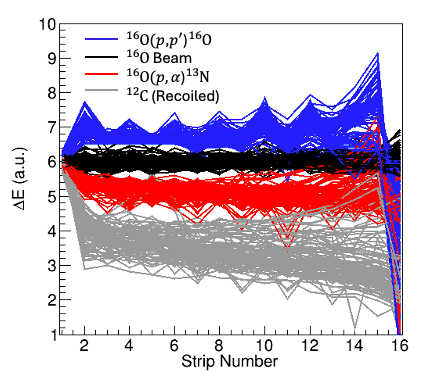}
\caption{Experimental $\Delta E$ traces recorded with MUSIC for events originating from strip~2. The $^{16}$O beam was normalised to 6\,a.u. (black). Elastic scattering events $^{16}$O$(p,p')$ appear with higher $\Delta E$ (blue). $(p,\alpha)$ events corresponding to $^{13}$N production are seen with lower $\Delta E$ (red). The lowest traces (grey) are attributed to recoiling $^{12}$C nuclei from collisions of $^{16}$O with $^{12}$C in the methane gas.}
\label{fig:Traces}
\end{figure}

Figure~\ref{fig:Traces} shows the experimental $\Delta E$ traces recorded using MUSIC for events initiating in strip~2. Energy loss signals are plotted versus strip number, with the $^{16}$O beam normalised to 6 a.u. Inelastic scattering events, primarily $(p,p')$, exhibit higher energy loss than the beam and appear as upward deviations. In contrast, $(p,\alpha)$ events associated with $^{13}$N production show a clear dip in energy loss, reaching 4.5–5.5\,a.u. Traces with even lower $\Delta E$ values, around 2–4.5\,a.u., are attributed to recoiling $^{12}$C nuclei produced in collisions of $^{16}$O with $^{12}$C atoms in the methane gas.

\begin{figure}
    \centering
    \includegraphics[width=1.0\linewidth]{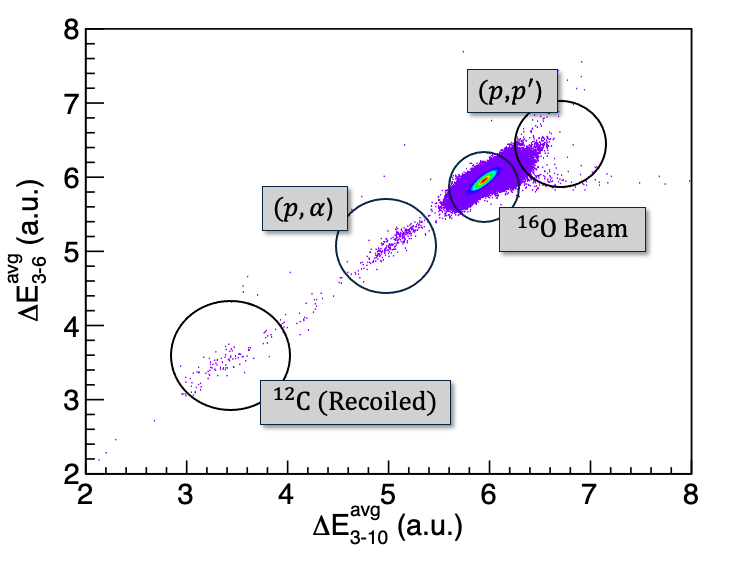}
    \caption{Two-dimensional energy loss ($\Delta E$) plot showing events assigned to strip~2 of the MUSIC detector. The vertical axis shows the average energy loss over anode strips 3–6, while the horizontal axis shows the average energy loss over strips 3–10. Distinct event populations corresponding to the unreacted ($^{16}$O) beam, inelastic scattering events $(p,p')$, $(p,\alpha)$ events producing $^{13}$N, and recoiling $^{12}$C nuclei are visible.}
    \label{fig:dEE}
\end{figure}

Event identification began by applying a gate on the first segmented strip (strip 1) to select primary beam events. Only events with energy loss within $E_{\mathrm{loss}}^{\mathrm{beam}} \pm 2\sigma$ were accepted, encompassing 95.4\% of the beam distribution while ensuring good separation from scattered particles.

A continuity condition suppressed irregular signals by requiring each event to produce a consistent $\Delta E$ trace across (8-10) consecutive strips.

The energy loss patterns were compared with simulations using ATIMA energy loss tables \citep{ATIMA}, which provided the expected $\Delta E$ values for $^{13}$N nuclei together with a $\pm10\%$ acceptance window. Based on these simulations, a minimum dip of 0.45\,a.u. in energy loss below the beam trace was required to identify a $(p,\alpha)$ event. In addition, to exclude beam fluctuations, only events with dips persisting across at least four consecutive strips were accepted. This continuity condition ensured that the observed signal corresponded to a real $^{13}$N particle rather than fluctuations in the $^{16}$O beam $\Delta E$.

The lower edge of the $\Delta E$ window for $^{13}$N, defined as 10\% below the predicted value, was used as a threshold: events below this limit were attributed to recoiling $^{12}$C nuclei. These were readily distinguishable, as their energy loss was substantially lower than that of $^{13}$N from the $(p,\alpha)$ reaction.

Figure~\ref{fig:dEE} shows a two-dimensional energy loss plot illustrating the clear separation between different particle types: beam particles $^{16}$O, inelastic scattering events $(p,p')$, $(p,\alpha)$ events producing $^{13}$N, and recoiling $^{12}$C nuclei from the methane gas. This separation demonstrates the effectiveness of the event identification criteria and provides additional event discrimination to the one-dimensional traces shown in Figure~\ref{fig:Traces}.

The $(p,\alpha)$ event count after all selection criteria represented the yield. The yield at each anode strip was used to calculate the cross section corresponding to the centre-of-mass energy for that strip (see Table~\ref{tab:cross section}). Systematic uncertainties arose mainly from two sources. The first was the uncertainty in beam intensity across detector strips; events within a $2\sigma$ window around the beam peak showed a 3\%  uncertainty due to  straggling. The second was differences in the $(p,\alpha)$ event counts obtained under different event selection (gating) conditions. These included 
changing the number of consecutive strips required to meet the continuity filter, and the depth of the energy loss dip relative to the beam trace (nominally set at 0.45\,a.u.). Varying these criteria tested the robustness of event identification and contributed to the estimate of systematic uncertainty.

Statistical uncertainties were evaluated using Poisson statistics \citep{919271} for high count rates, and the Feldman-Cousins method \citep{Feldman1998UnifiedSignals} for cases with fewer than 30 events.

The measured cross section values are summarised in Table~\ref{tab:cross section}.
\begin{table}[htbp]
    \caption{New experimental cross section data for the \reac\ reaction measured with the MUSIC detector.}
    \centering
    \begin{tabular}{ccccccc}
        \hline
        \hline
        Strip & $E_{\mathrm{cm}}$ & $\Delta E$ & counts & $\sigma$ & $\Delta\sigma_{\mathrm{sys}}$ & $\Delta\sigma_{\mathrm{stat}}$ \\
        number & (MeV) & (MeV) & \event & (mb) & (mb) & (mb) \\
        \hline
        6 & 5.78 & 0.15 & 1   & $\le$0.09 &  &  \\
        5 & 6.07 & 0.15 & 9   & 0.12 & 0.005 & $_{-0.08}^{+0.17}$ \\
        4 & 6.36 & 0.14 & 77  & 1.02 & 0.03  & 0.12 \\
        3 & 6.64 & 0.14 & 188 & 2.38 & 0.08  & 0.18 \\
        2 & 6.91 & 0.13 & 700 & 9.24 & 0.54  & 0.35 \\
        \hline
        \hline
    \end{tabular}
    \label{tab:cross section}
\end{table}

\section{Results}
\subsection{Updated cross section data}
\label{sec:Crossdata}

The new measurement of the \reac\ cross section, presented in Table~\ref{tab:cross section}, provides direct measurements in the low-energy region ($E_{\mathrm{cm}} = 5.8$–7.0~MeV). 

Although the experiment aimed to measure cross sections down to $E_{\mathrm{cm}} \approx 5.5$~MeV, very low event yields beyond strip~5 prevented reliable event identification. In strip~6, the number of $(p,\alpha)$ events was extremely low, and, for some choices of event identification conditions, consistent with zero counts. 
Additionally, a hardware issue in the readout system of strip~3 caused a $\sim$35\% reduction in the number of detected events, further complicating the statistical analysis for strip~6. Therefore the cross section for strip 6 is given as an upper limit.
Normally, the effective energy would be used, reflecting the energy dependence of the cross section across the strip. However, in this case the energy dependence includes unknown contributions from both narrow and broad resonances. For clarity, we therefore show the data point at the centre of the strip, to enable readers to implement their preferred energy dependence.

\begin{figure}[htbp]
    \centering
    \includegraphics[width=1\linewidth]{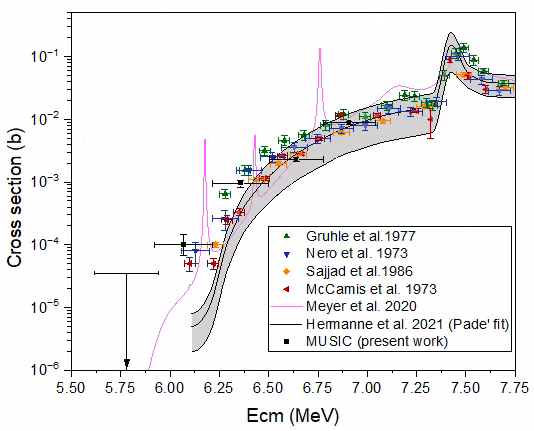}
    \caption{Measured cross section values of the \reac\ reaction obtained with the MUSIC detector (present work), compared with previously published data from \citet{gruhle1977reactions}, \citet{NERO}, \citet{SAJJAD1986Cyclotron}, and \citet{mccamis1973total}, after applying energy loss corrections. The results are also compared with the $\mathcal{R}$-matrix calculation of \citet{Meyer2020EvaluationGrains} and the Padé evaluation of \citet{Hermanne2021}. Error bars include both statistical and systematic uncertainties. The downward arrow at $E_{\mathrm{cm}} = 5.78$~MeV shows the upper limit from strip 6 analysis.}
    \label{fig:newexperimentaldata}
\end{figure}

The final cross section results from the present work are compared in Figure~\ref{fig:newexperimentaldata} with earlier measurements by \citet{NERO}, \citet{mccamis1973total}, \citet{gruhle1977reactions}, and \citet{SAJJAD1986Cyclotron}. For consistency, energy loss corrections were applied to the published cross section values from these datasets, using the values reported in the original studies or SRIM estimates \citep{ziegler2010srim} when unavailable. The figure also includes the cross section calculated from the inverse \reacE\ reaction, by \citet{Meyer2020EvaluationGrains} and converted using detailed balance, as well as the Padé evaluation of \citet{Hermanne2021}.

The cross section in the range  $E_{\mathrm{cm}} = 5.7$–7.0~MeV exhibits a steep decrease toward lower energies, consistent with the expected suppression due to the Coulomb barrier. The new MUSIC measurements follow this trend and show overall agreement with the shape of the excitation function observed in earlier studies.

Several observations arise from Figure~\ref{fig:newexperimentaldata}. First, the MUSIC results generally lie within the scatter of the other measurements, but our energy uncertainty limits the comparison. However, the narrow resonance predicted near $E_{\mathrm{cm}} \approx 6.75$~MeV in the $\mathcal{R}$-matrix calculation of \citet{Meyer2020EvaluationGrains} is not supported by the present data or by the measurements of \citet{NERO} and \citet{mccamis1973total}. This discrepancy suggests that the $\alpha$-width of the $E_{\mathrm{x}}=7.36$~MeV level in the compound nucleus of \F is likely much smaller than previously estimated. The relatively high cross section observed near $E_{\mathrm{cm}} \approx 6.0$–6.25~MeV in the MUSIC data indicates a resonance, as proposed by \citet{Meyer2020EvaluationGrains}. This enhancement is consistent with the earlier results of \citet{NERO}, providing additional support for a resonant structure in this region. 
Finally, the dataset of \citet{gruhle1977reactions} shows systematically higher cross section values than the present data and other measurements in this region, which may reflect experimental bias or uncorrected systematics.

\subsection{Updated Reaction Rate}
\label{sec:NewRate}

As discussed in Sec~\ref{sec:previous_data}, the Padé fit provides continuous coverage for the energy required to constrain the rate at temperatures of \(T = 3\)–4~GK. However, at lower energies (below \(E_{\mathrm{cm}} \approx 7.0\)~MeV), the reliability of the Padé fitting decreases due to significant discrepancies among the experimental datasets used in the fit, including those of \citet{gruhle1977reactions}, \citet{NERO}, \citet{mccamis1973total}, \citet{SAJJAD1986Cyclotron}, and \citet{whitehead1958activation}. These inconsistencies introduce substantial uncertainty of 65\%. To improve the description of the cross section in this energy region and reduce the associated uncertainty, the present MUSIC measurements were adopted for \(E_{\mathrm{cm}} < 7.0\)~MeV. However, the data point at \(E_{\mathrm{cm}} = 5.78 \pm 0.15\)~MeV, shown as an upper limit in Fig.~\ref{fig:newexperimentaldata}, was not included in the present rate analysis. This point lies outside the energy range that contributes significantly to the reaction rate at temperatures of \(T = 3\)–4~GK and therefore has a negligible effect on the calculated rate. Consequently, it was excluded from the reaction rate calculation presented in this work.
Accordingly, the adopted cross section for the reaction rate calculation combines the MUSIC data for \(E_{\mathrm{cm}}<7.0\)~MeV with the Padé fitting for \(E_{\mathrm{cm}}>7.0\)~MeV.

The final adopted dataset was used as input to the \texttt{Exp2Rate} code \citep{rauscher2000astrophysical} to compute the reaction rate with uncertainties over $T = 2$–5\,GK. The calculated rates are summarised in Table~\ref{tab:rateOf16O}.
\begin{table}[htbp]
    \centering
    \caption{Calculated thermonuclear reaction rates for the \reac\ reaction derived using the \texttt{EXP2Rate} code with the results from the present work and the adopted cross section.}
    \label{tab:reaction_rates}
    \begin{tabular}{cccc}
\hline\hline
 $T_9$ & Low & Recommended & High \\
(GK) & (cm$^{3}$\,s$^{-1}$\,mol$^{-1}$) & (cm$^{3}$\,s$^{-1}$\,mol$^{-1}$) & (cm$^{3}$\,s$^{-1}$\,mol$^{-1}$) \\
\hline
 2.5 & $8.85\times10^{-6}$ & $1.20\times10^{-5}$ & $1.62\times10^{-5}$ \\
3.0 & $1.21\times10^{-3}$ & $1.67\times10^{-3}$ & $2.30\times10^{-3}$ \\
 3.5 & $4.19\times10^{-2}$ & $5.95\times10^{-2}$ & $8.46\times10^{-2}$ \\
 4.0 & $6.07\times10^{-1}$ & $8.88\times10^{-1}$ & $1.30\times10^{0}$ \\
4.5 & $4.89\times10^{0}$ & $7.32\times10^{0}$ & $1.09\times10^{1}$ \\
 5.0 & $2.60\times10^{1}$ & $3.96\times10^{1}$ & $6.03\times10^{1}$ \\

    \hline\hline
    \label{tab:rateOf16O}
    \end{tabular}
\end{table}

\begin{figure}[htbp]
    \centering
    \includegraphics[width=1\linewidth]{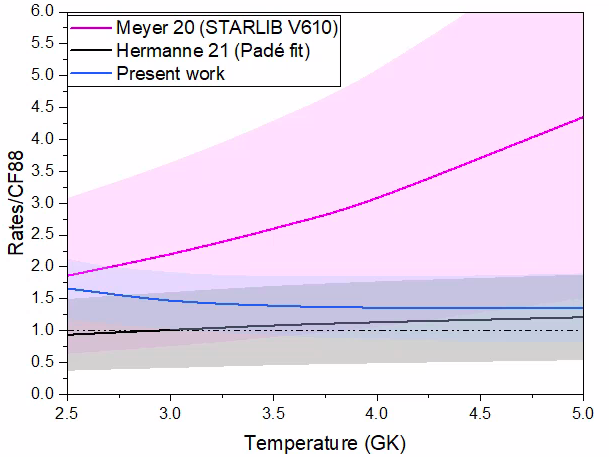}
    \caption{Ratio of the new reaction rate (blue line) to the CF88 rate (black dotted dashed line) as a function of temperature ($T = 2.5$–5\,GK), relevant for explosive oxygen burning in Type Ia supernovae. 
    The comparison includes the rate of \citet{Meyer2020EvaluationGrains} as provided in the latest STARLIB \citet{sallaska2013starlib}  and the IAEA evaluation of \citet{Hermanne2021}. Shaded bands indicate the one sigma uncertainty of each dataset.}
    \label{fig:rateSNIa}
\end{figure}
\figf{fig:rateSNIa} shows the calculated reaction rate relative to the CF88 rate. The comparison includes the rate of \citet{Meyer2020EvaluationGrains} as provided in the latest STARLIB \citet{sallaska2013starlib} and the IAEA Padé fit of the \reac\ cross section \citep{Hermanne2021} is also shown.

The present rate exhibits a different temperature dependence from the rate of \citet{Meyer2020EvaluationGrains}, with the discrepancy becoming more pronounced above $T_9 \approx 3$ (Figure~\ref{fig:rateSNIa}). 
This behaviour is expected, since \citet{Meyer2020EvaluationGrains} normalised their rate above 1.4 GK using the \reac\ rate from STARLIB, which is based on Hauser–Feshbach (HF) models that tend to overpredict cross sections for light, low-level-density nuclei~\citep{sallaska2013starlib}.
The comparison between present work and the rate from Padé fit (\citet{Hermanne2021}) shows how the low energy data changed the rate behaviour. The present rate ratio is  higher below $T_9 \approx 4.5$, primarily due to differences in the cross section values within the $E_{\mathrm{cm}} = 5.8$–6.25~MeV range, where the MUSIC data exceed the Hermanne evaluation by up to a factor of 3–10 (Figure~\ref{fig:newexperimentaldata}).

The presented reaction rate only considered reactions on the ground state of $^{16}$O. However, a possible contribution from thermally populated excited states of $^{16}$O could impact the rate at the relevant temperatures. Although the lowest lying excited states in $^{16}$O are at relatively high excitation energies of around 6~MeV, the low probability for thermally populating them is exactly balanced by the high probability of available protons with the lower energy required to populate these compound states (10-100s keV rather than 6-7~MeV). The reaction rate on thermally populated states therefore depends on the relevant proton widths ($\Gamma_{p1}$, $\Gamma_{p2}$, etc). The emission/capture of these lower energy protons ($p_1$, $p_2$, etc) is inhibited by the Coulomb barrier while the ground state channel is not. Where the $p_1$ channel is measured, for states above 7.7~MeV, it is at most 25\% 
of the $p_0$ channel, and this factor will decrease rapidly with decreasing excitation energy. Therefore, while it is important to understand the contribution of thermally excited \Ox states to fully constrain the reaction rate, the overall conclusions of this work remain the same.

\subsection{Astrophysical Implications for SNe Ia}

\begin{table*}[!t]
\centering
\caption{
Sensitivity of the Ca/S and Ar/S ejecta mass ratios to variations in the key nuclear reaction rates governing explosive oxygen burning. The listed Ca/S and Ar/S values correspond to the ranges obtained when varying the progenitor metallicity from $Z = 0.0014$ to $Z = 0.042$ (0.01--3\,$Z_\odot$). Rate multipliers indicate enhancement or reduction factors relative to the nominal rate. For $(p,\alpha)$ and $(\gamma,\alpha)$ reactions, the corresponding reverse rates were scaled by the same factor.}
\label{tab:RateSensitivity}
\begin{tabular}{ccccccc}
\hline\hline
Model &
$^{16}$O$(p,\alpha)^{13}$N &
$^{12}$C+$^{16}$O &
$^{16}$O+$^{16}$O &
$^{16}$O$(\gamma,\alpha)^{12}$C &
Ca/S &
Ar/S \\
\hline
S & $\times1$ & $\times1$ & $\times1$ & $\times1$ & 0.19--0.24 & 0.21--0.23 \\
A & $\times2$ & $\times1$ & $\times1$ & $\times1$ & 0.19--0.27 & 0.21--0.24 \\
B & $\times2$ & $\times0.3$ & $\times1$ & $\times1$ & 0.22--0.32 & 0.23--0.27 \\
C & $\times2$ & $\times1$ & $\times0.3$ & $\times1$ & 0.20--0.28 & 0.21--0.25 \\
D & $\times2$ & $\times0.5$ & $\times0.5$ & $\times1$ & 0.21--0.31 & 0.22--0.26 \\
E & $\times2$ & $\times0.7$ & $\times1$ & $\times3$ & 0.22--0.30 & 0.23--0.26 \\
\hline\hline
\end{tabular}
\end{table*}

\begin{figure*}[!t]
    \centering
    \begin{subfigure}[t]{0.47\textwidth}
        \centering
        \includegraphics[width=\linewidth]{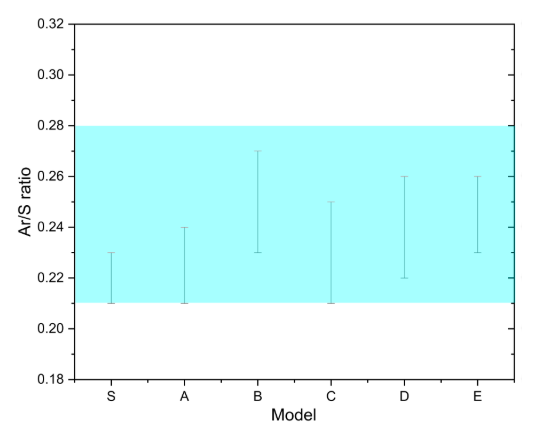}
        \caption{Ar/S ejecta mass ratio.}
        \label{fig:ArS}
    \end{subfigure}
    \hfill
    \begin{subfigure}[t]{0.47\textwidth}
        \centering
        \includegraphics[width=\linewidth]{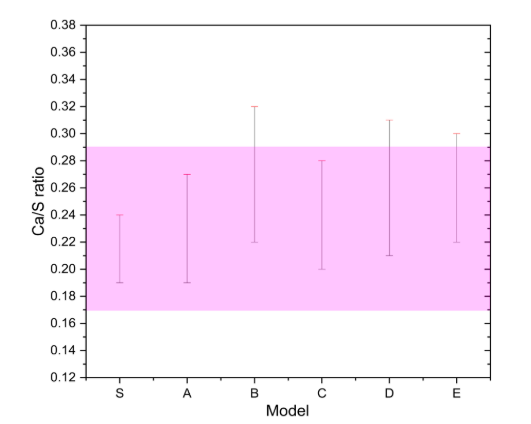}
        \caption{Ca/S ejecta mass ratio.}
        \label{fig:CaS}
    \end{subfigure}
    \caption{
Predicted ranges of the Ar/S (left) and Ca/S (right) ejecta mass ratios for the different reaction-rate variation models listed in Table~\ref{tab:RateSensitivity}. Vertical bars indicate the variation of the ratios obtained when the progenitor metallicity is varied from $Z = 0.0014$ to $Z = 0.042$ (0.01--3\,$Z_\odot$). The shaded bands represent the observational ranges inferred for extreme supernova remnants, Ar/S $\simeq$ 0.21--0.28 and Ca/S $\simeq$ 0.17--0.29, as reported by~\citet{martinez2017observational}.}
    \label{fig:BravoCas}
\end{figure*}
Previous work by \citet{bravo201916o} emphasised the role of the \reac\ reaction in regulating the $\alpha$-particle abundance during explosive oxygen burning in Type Ia supernovae. Their study suggested that the metallicity dependence of the calcium-to-sulphur Ca/S production ratio could be reproduced if the \reac\ reaction rate were enhanced by up to a factor of seven relative to CF88. In contrast, the reaction rate derived in the present work shows only a modest enhancement, by a factor of $\sim$1.5-2 in the $T = 3$–4\,GK range. This falls well below the factor of seven proposed by \citet{bravo201916o} as necessary to reproduce the full spread of observed Ca/S ratios. Nevertheless, the new rate lies above the lower limit of $0.5 \times$ CF88 suggested by \citet{bravo201916o}, supporting the conclusion that \reac\ does contribute to the sensitivity of Ca/S to progenitor metallicity.

To assess the impact of the new \reac rate constraints on Ca/S and Ar/S production, SNe~Ia simulations were carried out at metallicities of $Z = 0.01\,Z_\odot$ and $Z = 3\,Z_\odot$, adopting standard values for all oxygen destruction reaction rates as suggested by~\citet{2019MNRAS.482.4346B}, specifically \reac, $^{12}$C+$^{16}$O, $^{16}$O+$^{16}$O, and $^{16}$O$(\gamma,\alpha)^{12}$C. Then, the \reac rate was enhanced by a factor of two, consistent with the upper bound of the new experimental constraint reported in this work, in order to isolate its direct impact on Ca/S and Ar/S production.

In addition, combined variations of the other oxygen destruction channels were explored by fixing the factor of two enhancement to \reac rate with independent modifications of the $^{12}$C+$^{16}$O, $^{16}$O+$^{16}$O, and $^{16}$O$(\gamma,\alpha)^{12}$C reaction rates. The resulting Ca/S and Ar/S ejecta mass ratios were compared with the observational ranges inferred for supernova remnants by \citet{martinez2017observational}. The model rate modifications and ratios predictions are summarised in Table~\ref{tab:RateSensitivity} and illustrated in Fig.~\ref{fig:BravoCas}.

As shown in Fig.~\ref{fig:BravoCas}, enhancing only the \reac rate (model~A) significantly improves the agreement with the observed Ca/S ratios relative to the standard rate model (Fig.~\ref{fig:CaS}), while producing only a modest improvement to the Ar/S production (Fig.~\ref{fig:ArS}). This demonstrates that the \reac reaction plays a key role in regulating calcium production during explosive oxygen burning in agreement with~\citet{Woosley1972Astrophysical16Opa} and~\citet{bravo201916o}, but that variations of this rate alone are insufficient to reproduce the full range of both Ca/S and Ar/S observed in supernova remnants. Consequently, additional variations in other oxygen destruction channels must also be considered.

Recently, the issue of mass ratio measurements in X-ray bright Type Ia SNRs has been revisited by Holland-Ashford et al., who re-analyzed the Suzaku observations of Kepler~\citep{Holland2023} and Tycho~\citep{Holland2025}. These authors took into account a number of effects, including effective-area calibration uncertainties in the Suzaku detectors, and uncertainties in the emitting volumes assumed for each species, and produced the most accurate measurements of mass ratios in the ejecta of these SNRs to date. For the Ca/S mass ratio, they measure values of 0.333 ± 0.080 for Kepler and 0.40 ± 0.15 for Tycho, which are slightly higher than the values found in \citet{martinez2017observational}, but statistically consistent given the higher, more realistic, uncertainties that they quote. These results reinforce the finding that the yields in most Type Ia SN explosion models underpredict the Ca/S mass ratios observed in SNRs. 

While none of the models fully reproduces the observed ranges, the new experimental constraint on the \reac rate 
excludes this reaction as a sole solution to the discrepancies with observations, and underscores the importance of constraining other oxygen destruction channels, given the sensitivity of Ca/S and Ar/S ratios to these processes. In addition, it should be noted that some of the  discrepancies may also result from observational uncertainties, as several of the X-ray–derived abundance ratios reported by \citet{martinez2017observational} have uncertainties of up to $\sim$0.02–0.03 (1$\sigma$).

\label{sec:RateImpact}

\section{Conclusion}
\label{sec:con}
In this work, we have derived a new reaction rate for \reac, experimentally constrained by a direct measurement of the reaction cross section with the MUSIC active-target detector. The updated rate is enhanced by around $\sim$1.5 times CF88 at $T=3$–4\,GK, significantly below the factor-of-seven increase proposed in previous modelling studies. This modest enhancement indicates that while the \reac\ rate contributes to the sensitivity of Ca/S to progenitor metallicity in Type~Ia supernovae, it cannot fully account for the observed variations in Ca/S and Ar/S ratios in supernova ejecta.

A reduction in other oxygen destruction reaction rates is therefore also required. 
However, new measurements by \citet{Fang} constrain the $^{12}$C+$^{16}$O cross section to, typically, within $\sim$20\%. Their evaluation suggests the rate between 3 and 5 GK deviates by less than 20\% from CF88.  
Confirmation of this result is needed, particularly given the up to a factor of two discrepancy between earlier measurements by \citet{Patterson} and \citet{Christensen}.
Further improvements in constraining the \reac\ reaction rate at lower energies would clarify the extent to which other reaction rate variations are required. More broadly, our findings indicate that current nucleosynthesis models of Type~Ia supernovae may need revision in order to bring theoretical predictions into agreement with the observed abundance ratios.

\section{Acknowledgements}
This research used resources of ANL’s ATLAS facility, which is a DOE Office of Science User Facility. This material is based upon work supported by the U.S. Department of Energy, Office of Science, Office of Nuclear Physics, under contract number DE-AC02-06CH11357. AML and CD thank the UK Science Technology and Facilities Council (STFC) for their support through grant ST/Y000285/1. This material is based partly upon work supported by the U.S.Department of Energy, Office of Science, Office of Nuclear Physics,
under Contract Nos. DE-FG02-97ER41033
and DE-FG02-97ER41042. E.B. acknowledges partial support from the Spanish grant PID2021-123110NB-100 funded by MICIU/AEI/10.13039/501100011033 and by FEDER, UE. The authors would also like to thank Peter Mohr for useful discussions.

\bibliography{References}{}

\bibliographystyle{aasjournalv7}

\end{document}